\documentclass[%
 aip,
pop,
 amsmath,amssymb,
 reprint,%
]{revtex4-1}

\usepackage{amsmath}
\usepackage{graphicx}
\usepackage{subcaption} 
\usepackage{dcolumn}
\usepackage{bm}
\usepackage{hyperref}
\usepackage[mathlines]{lineno}
\usepackage{dirtytalk}
\usepackage{diffcoeff}  
\usepackage[inline]{enumitem}
\usepackage{mathtools}
\usepackage{mathrsfs}  
\usepackage{siunitx}

\usepackage[utf8]{inputenc}
\usepackage[T1]{fontenc}
\usepackage{mathptmx}
\usepackage{etoolbox}
\usepackage{tabto}
\usepackage{nomencl}
\usepackage{ifthen}
\usepackage{changes}

\renewcommand{\nomgroup}[1]{%
  \ifthenelse{\equal{#1}{E}}{\item[\textbf{The Compressible Liquid Regime}]}{%
  \ifthenelse{\equal{#1}{D}}{\item[\textbf{Criterion for the Reversal of the Bjerknes Force}]}{%
  \ifthenelse{\equal{#1}{C}}{\item[\textbf{The Acoustic Cycle}]}{%
  \ifthenelse{\equal{#1}{B}}{\item[\textbf{The Mapping in the Incompressible Regime}]}
  {\item[\textbf{Introduction}]}
  }}}}
\makenomenclature

\makeatletter
\def\@email#1#2{%
 \endgroup
 \patchcmd{\titleblock@produce}
  {\frontmatter@RRAPformat}
  {\frontmatter@RRAPformat{\produce@RRAP{*#1\href{mailto:#2}{#2}}}\frontmatter@RRAPformat}
  {}{}
}%
\makeatother

\hypersetup{
    colorlinks=true,
    linkcolor=blue,
    filecolor=magenta,      
    urlcolor=cyan,
    pdftitle={Shimon Stern 2024},
    pdfpagemode=FullScreen,
    }
\begin{document}

\preprint{AIP/123-QED}

\title{Mapping Driven Oscillations in the Size of a Bubble to the Dynamics of a Newtonian Particle in a Potential}

\author{Uri Shimon}
\email{u.shimon@weizmann.ac.il}
\author{Ady Stern}%
\affiliation{%
	Department of Condensed Matter Physics, Weizmann Institute of Science, Rehovot 76100, Israel
}%


\date{\today}

\begin{abstract}
The non-linear dynamics of driven oscillations in the size of a spherical bubble are mapped to the dynamics of a Newtonian \textit{particle in a potential} within the incompressible liquid regime. The compressible liquid regime, which is important during the bubble’s sonic collapse, is approached adiabatically. This new framework naturally distinguishes between the two time scales involved in the non-linear oscillations of a bubble. It also explains the experimentally observed sharp rebound of the bubble upon collapse. Guided by this new vantage point, we develop analytical approximations for several key aspects of bubble motion. First, we formulate a tensile strength law that integrates the bubble's ideal gas behavior with a general polytropic index. Next, we establish a straightforward physical criterion for Bjerknes force reversal, governed by the driving pressure, ambient pressure and tensile strength. Finally, we derive an acoustic energy dissipation formula for the bubble's sonic collapse, dependent solely on the bubble's collapse radii and velocity. 
\end{abstract}

\maketitle

\nomenclature[Aa]{$Q$}{Bubble radius raised to the power $5/2$}
\nomenclature[Aa]{$R$}{Bubble radius}
\nomenclature[Aa]{$t$}{Time}
\nomenclature[Aa]{$p_{GE}$}{Ideal gas pressure when $R=R_{E0}$}
\nomenclature[Aa]{$p_V$}{Vapor pressure}
\nomenclature[Aa]{$p_0$}{Ambient pressure}
\nomenclature[Aa]{$p_d$}{Driving pressure}
\nomenclature[Ba]{$\bar{p}_d$}{Driving pressure amplitude}
\nomenclature[Ab]{$\rho_L$}{Density of the liquid}
\nomenclature[Da]{$R_C$}{Linear Bjerknes force critical radius}
\nomenclature[Db]{$\Phi$}{Bjerknes force time correlator}
\nomenclature[Ab]{$\nu_L$}{Kinematic viscosity}
\nomenclature[Aa]{$R_{E0}$}{Equilibrium radius in the absent of driving}
\nomenclature[Ba]{$R_{E}$}{Instantaneous equilibrium radius}
\nomenclature[Ba]{$Q_{E}$}{$R_E^\frac{5}{2}$}
\nomenclature[Ba]{$Q_{E0}$}{$R_{E0}^\frac{5}{2}$}
\nomenclature[Aa]{$S$}{Surface tension}
\nomenclature[Aa]{$k$}{Polytropic index}
\nomenclature[Aa]{$M$}{Mach number, $\dot{R} / c$}
\nomenclature[Cb]{$\tau_\star$}{Instance at which $\Delta p=\Delta p_c$}
\nomenclature[Cb]{$\Delta \tau$}{$\tau-\tau_\star$}
\nomenclature[Cb]{$\tau$}{Dimensionless time $\tau=\omega_dt$}
\nomenclature[Ea]{$c$}{Speed of sound in the liquid}
\nomenclature[Ab]{$\Delta p_c$}{Tensile strength: the tension beyond which significant bubble expansion is anticipated}
\nomenclature[Ab]{$\Delta p$}{Tension, $p_V-p_0-p_d(t)$}
\nomenclature[Ab]{$\omega_d$}{Driving frequency}
\nomenclature[Eb]{$\lambda$}{Tunable parameter between 0 and 1}
\nomenclature[Ba]{$U_k$}{Instantaneous potential in the incompressible liquid regime}
\nomenclature[Ea]{$D$}{Effective dimension}
\nomenclature[Eb]{$\alpha$}{Radius exponent in the compressible regime}
\nomenclature[Ea]{$X$}{Bubble radius raised to the power $\alpha$}
\nomenclature[Cb]{$\tau_{max}$}{Instance at which the bubble reaches its maximal radius}
\nomenclature[Da]{$F_B$}{Primary Bjerkness force}
\nomenclature[Ea]{$h_B$}{Enthalpy of the bubble}
\nomenclature[Ea]{$R_{\text{max}}$}{Maximal radius}
\nomenclature[Ea]{$R_{\text{min}}$}{Minimal radius}
\nomenclature[Ea]{$E$}{Energy of the bubble}
\nomenclature[Ea]{$E_{\text{acoustic}}$}{Generated acoustic energy upon bubble collapse}

\printnomenclature
\section{\label{sec:introduction}Introduction}
Spherical micro bubble collapse can lead to immense energy-per-particle focusing of almost 12 orders of magnitude, leading to light emission known as \textit{sonoluminescence} \cite{Putterman2000,Matula1999,Weninger1997,Brennan2016,Lauterborn2007,Putterman1995,Kamath1993,Lfstedt1993,Brenner2002}. The shock wave generated in the process has associated pressures on the order of GPa \cite{Akhatov2001,Pecha2000,Supponen2017}. Such pressures can wear metallic surfaces and are  a major challenge in designing machinery that involves liquids \cite{Arndt1981,Vogel1989,Escaler2006,Rambod1999,NASA1971,d’Agostino2017,Li2021}. The damaging power can be used in various novel medical applications \cite{Brennen2015,Field1991,Sass1991,BiasioriPoulanges2024,Yu2004}  and for water treatment \cite{Dular2016,Song2022,Song2004}. It is also used by the Mantis shrimp in the killing of its prey\cite{Patek2005,Versluis2000}.

\begin{figure}[t]
    \centering
    \includegraphics[width=1\linewidth]{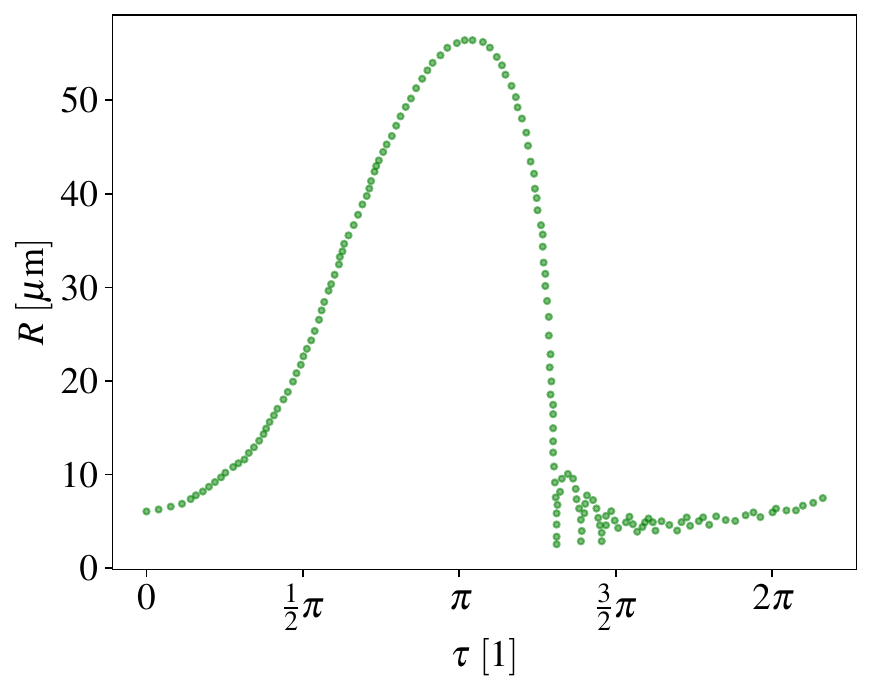}
    \caption{The bubble radius versus dimensionless time $\tau=\omega_dt$, where $\omega_d$ is the driving frequency. Reproduced from Putterman and Weninger, Annu. Rev. Fluid. 32, 445 (2000) \cite{Putterman2000}; licensed under a CC BY license.}
    \label{fig:experimental}
\end{figure}

In the presence of an oscillating acoustic field, the bubble collapse can be made periodic, encompassing two distinct time scales: the \textit{slow} driving frequency measured in kHz, and the \textit{fast} MHz oscillations that arise post the initial collapse of the bubble (referred to as "after bounces") \cite{Putterman2000,Matula1999,Weninger1997,Brennan2016}, see Fig. \ref{fig:experimental}. It also spans a broad range of length scales, from 0.1 to 10 $\SI{}{\micro\metre}$. 

The dynamics of a spherical bubble with radius $R(t)$ in the incompressible liquid regime are described by the well-known Rayleigh-Plesset equation \cite{Basset1888,Rayleigh1917,Plesset1949,Brennen2013,Acoustic1994}:
\begin{align}
	R\ddot{R}+\frac{3}{2}\dot{R}^2=\frac{p_{GE}\left(\frac{R_{E0}}{R}\right)^{3k}+p_V-p_0-p_d(t)}{\rho_L} 
 \nonumber \\
 -\frac{4\nu_L}{R}{\dot{R}} -\frac{2S}{\rho_LR}
	\label{eq:rayleigh_plesset}
\end{align}
The bubble is treated as an ideal gas going through a polytropic process with index $k$, $p_{GE}$ being the ideal gas pressure when $R=R_{E0}$, where $R_{E0}$ is the equilibrium radius in the absent of driving (will be refereed to as the ambient equilibrium radius). $S$ is the surface tension, $\nu_L$ the liquid's kinematic viscosity and $p_V,p_0$ are the vapor and ambient pressures, respectively. The driving pressure, $p_d(t)$, is the oscillatory pressure associated with the acoustic wave, characterized by the frequency $\omega_d$. We will refer to the pressure difference $\Delta p(t)\equiv p_V-p_0-p_d(t)$ as the tension. The tensile strength is the critical tension, $\Delta p_c$, at which significant bubble expansion takes place \cite{Brennen2013}.

For an oscillating bubble in sonoluminescence conditions, the incompressibility of the liquid is a sensible assumption throughout the entire motion of the bubble, except during the bubble's sonic collapse \cite{Prosperetti1986,Putterman2000}. Perturbation theory in the bubble's Mach number $\left(M\equiv\frac{\dot{R}}{c}\right)$ can be used to extend (\ref{eq:rayleigh_plesset}) to the compressible liquid regime\cite{Prosperetti1986}. We will review this extension in Section \ref{sec:the_compressible_regime}, which addresses the compressible liquid regime.

The primary Bjerknes force is a translational force acting on an acoustically driven bubble. It is defined as the time-averaged product of the pressure gradient and the bubble's volume over an acoustic cycle, \(\langle \nabla p R^3 \rangle\) \cite{Bjerknes1906,Leighton1990,Crum1975,Louisnard2008}. In the presence of a \textit{weak} standing acoustic wave, bubbles with a small ambient equilibrium radius are drawn towards the pressure anti-nodes, while those with a large ambient equilibrium radius are drawn towards the nodes. Consequently, the Bjerknes force dictates the effective pressure a bubble experiences and plays a key role in defining the parameter space for stable sonoluminescence \cite{Akhatov1997,Louisnard2008}.

\begin{figure*}[t] 
    \centering
    \begin{subfigure}[t]{0.32\textwidth}
        \centering
        \includegraphics[width=\linewidth]{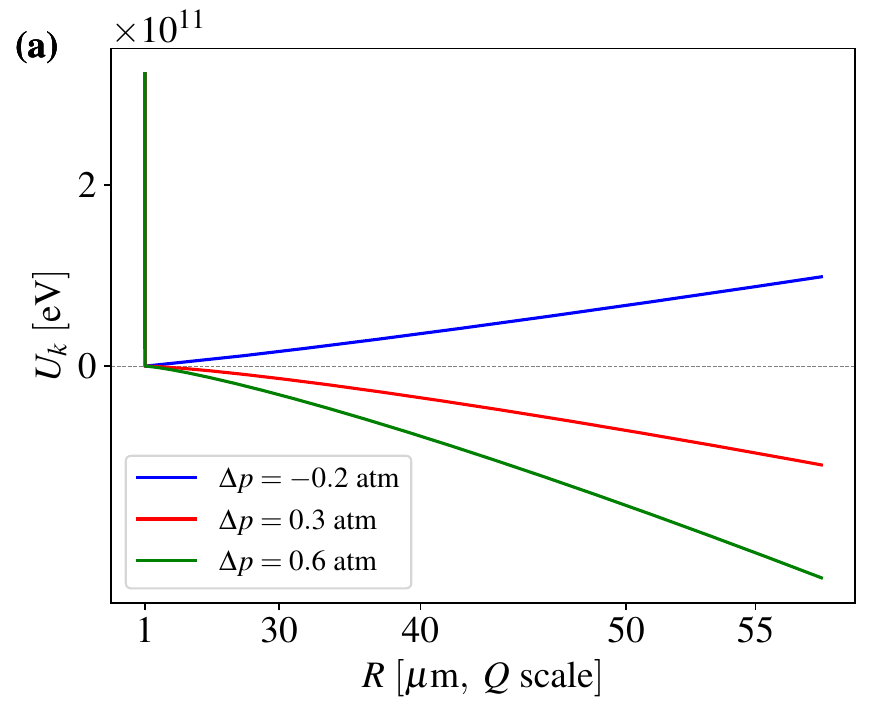} 
        \captionsetup[sub]{position=top}
    \end{subfigure}
    \begin{subfigure}[t]{0.32\textwidth}
        \centering
        \includegraphics[width=\linewidth]{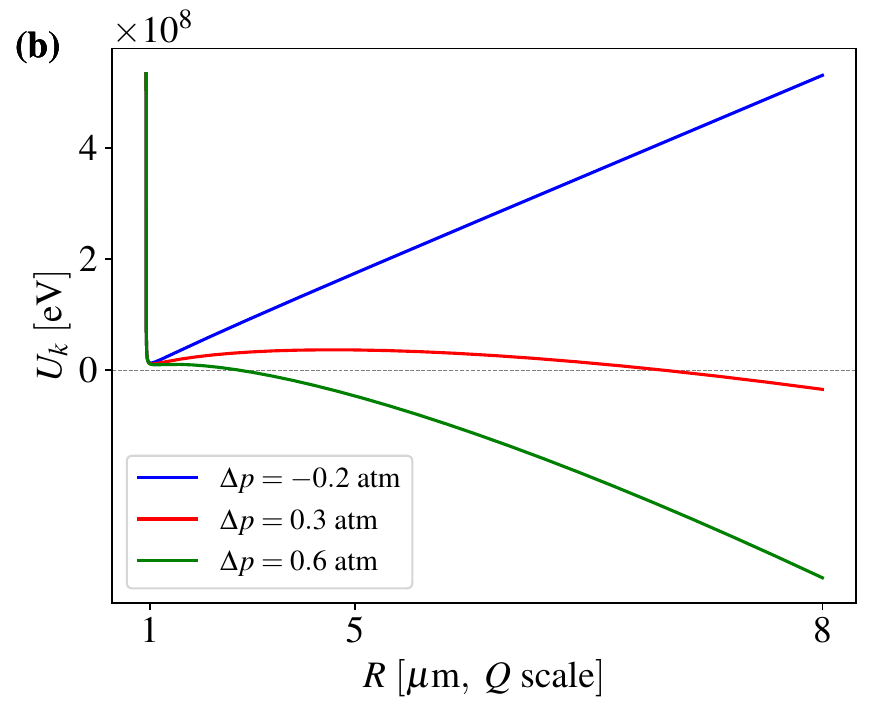} 
    \end{subfigure}
    \begin{subfigure}[t]{0.32\textwidth}
        \centering
        \includegraphics[width=\linewidth]{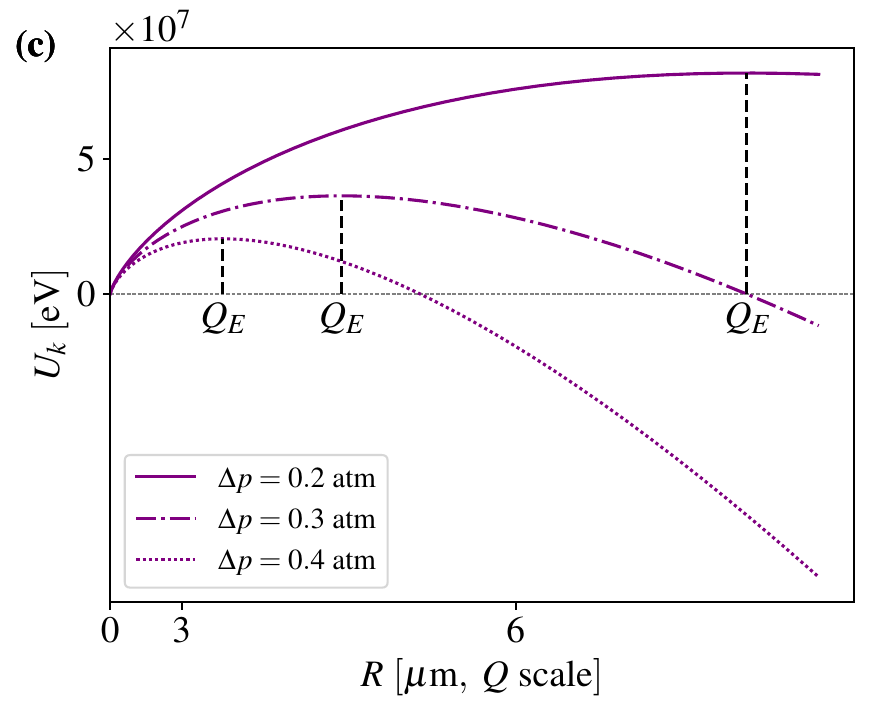} 
    \end{subfigure}
    \caption{The bubble's instantaneous potential, $U_k$, versus its radius, $R$, in the incompressible liquid regime $\bigl($Eq. (\ref{eq:a_particle_in_a_potential}), $Q_{E0}=\SI{1}{\micro\metre\tothe{\frac{5}{2}}}$, $S=\SI{0.0728}{\newton\over\metre}\bigr)$ plotted in $Q=R^\frac{5}{2}$ scale. \textbf{(a)}-\textbf{(b)} The long and short range structure of the potential for $k=\frac{5}{3}$. For negative tension (the blue curve), a single stable equilibrium radius exists and the potential is binding. For positive tension (red and green curves), the potential is non-binding in the long range. However, if the tension is lower than the tensile strength (the red curve), a meta-stable equilibrium radius exists, with a locally binding potential around it. \textbf{(c)}  The potential for $k<\frac{1}{3}$ and varying tension. The unstable equilibrium point, $Q_E(\Delta p)$, approaches zero as the tension is increased.}
    \label{fig:bubble_potential}
\end{figure*}

In this work, we transform the non-linear Rayleigh-Plesset equation (Eq. (\ref{eq:rayleigh_plesset})) to a linear equation of motion of a Newtonian particle in a potential. Our starting point is the transformation $Q=R^\frac{5}{2}$, introduced first in a more limited context by Childs \cite{Childs1973}. The transformation for the incompressible liquid regime is presented in Section \ref{sec:the_potential_in_the_incompressible}. This novel mapping naturally distinguishes between the two time scales of the driven motion. The slow time scale is the acoustic period, while the fast time scale is the one associated with oscillations of the Newtonian particle within its potential. The resulted potential, drawn in Fig. \ref{fig:bubble_potential}, manifests a very sharp wall at $Q\rightarrow0$ for polytropic index, $k>\frac{1}{3}$. This sharp wall leads to very high acceleration at the bubble's rebound from its collapse, which is observed experimentally\cite{Putterman2000,Matula1999,Weninger1997} (see Fig. \ref{fig:experimental}). It also indicates that such a rebound is not expected for $k<\frac{1}{3}$. Section \ref{sec:the_potential_in_the_incompressible} is concluded with the calculation of the critical tension, $\Delta p_c\equiv(p_V-p_0-p_d)_c$, at which significant bubble expansion is anticipated, including its dependence on $S$ and $k$ \cite{Brennen2013}. This critical tension will be referred to as the tensile strength. 

In Section \ref{sec:the-acoustic-cycle} we describe one acoustic cycle of the bubble by dividing it into segments. Section \ref{sec:bjerknes_force_reversal} is dedicated to the derivation of a physical criterion for the reversal of the primary Bjerknes force. The criterion relies exclusively on the driving pressure, ambient pressure, and tensile strength. 

Finally, in Section \ref{sec:the_compressible_regime}, the mapping is extended in an approximated way to the compressible liquid regime. An approximately conserved quantity (up to viscosity losses) is then found by assuming slowly varying bubble's Mach number $\left(M\equiv\frac{\dot{R}}{c}\right)$. The conserved quantity is used to calculate the acoustic energy dissipation during collapse, expressed as a function of the bubble's Mach number, as well as its initial, final, and equilibrium radii.
\section{The Mapping in the Incompressible Regime\label{sec:the_potential_in_the_incompressible}}
To map (\ref{eq:rayleigh_plesset}) to the dynamics of a Newtonian particle in a potential, we apply the transformation $Q=R^\frac{5}{2}$. This transforms Eq. (\ref{eq:rayleigh_plesset}) to an equation of motion of a non-relativistic particle subject to a time-dependent potential and a spatially varying frictional force:
    \begin{align}
	&\rho_L\ddot{Q}=-\frac{d}{dQ}U_k(Q)-4\rho_L\nu_L\frac{\dot{Q}}{Q^\frac{4}{5}}
        \label{eq:a_particle_in_a_potential}
    \end{align}
where
    \begin{align}
    U_k&=
	\frac{25}{12}\left(\frac{p_{GE}}{(k-1)} \left(\frac{Q_{E0}}{Q}\right)^{\frac{6k}{5}}
	+ p_0+p_d(t)-p_V\right)Q^\frac{6}{5}
 \nonumber \\
    &+\frac{25}{4}SQ^{\frac{4}{5}}
    \label{eq:the_incompressible_potential}
    \end{align} wherein $Q_{E0}=R_{E0}^\frac{5}{2}$ denotes the ambient equilibrium point.
For $k=1$, the polytropic term changes to $-\frac{5}{2}p_{GE} Q_{E0}^{\frac{6}{5}}\log{\left(\frac{Q}{Q_{E0}}\right)}$. The resulting dynamics may take several qualitatively different forms, depending on the number and type of extremal points of $U_k(Q)$. This number is determined by the number of \textit{positive} solutions of:
\begin{align}
	\partial_QU_k(Q)=0 \iff p_{GE}Q_{E0}^{\frac{6k}{5}}+\Delta pQ_E^\frac{6k}{5}=2SQ_E^\frac{6k-2}{5}
    \label{eq:equilibrium_general_k}
\end{align}
We denote these solutions by $Q_E$. 

The tensile strength, $\Delta p_c \ge 0$, can now be calculated, and to that end we now consider a positive tension. For $k<\frac{1}{3}$, the potential has a single unstable equilibrium point which approaches $0$ as the tension is increased (Fig. \ref{fig:bubble_potential}c). The bubble expands when its size is larger than its equilibrium point, which happens when the tension overcomes the surface tension:
\begin{align}
    \Delta p_c = \frac{2S}{Q_{E0}^\frac{2}{5}}-p_{GE} \mathrm{\  } \left(k<\frac{1}{3}\right)
\end{align}
The tensile strength is then independent of $k$. 

If $k>\frac{1}{3}$, for strong positive tension there will be no equilibrium point in the potential, as seen in the green curve in Fig. \ref{fig:bubble_potential}a-b. In contrast, for weak positive tension there will be one maximum and one minimum point to the potential, as seen in the red curve in Fig. \ref{fig:bubble_potential}b. The system is assumed to start in the weak tension regime, at the minimum point of the potential. The second derivative of the potential does not change sign in the strong tension case, but does go through zero in the weak tension case. Its examination, then, allows us to extract the tensile strength:
\begin{align}
    \Delta p_c = \frac{2(3k-1)}{3k}\left(\frac{2S^{3k}}{3kp_{GE}Q_{E0}^\frac{6k}{5}}\right)^\frac{1}{3k-1}
    \mathrm{\ } \left(k>\frac{1}{3}\right)
    \label{eq:nucleation-formula}
\end{align} The tensile strength is then $k$-dependent for $k>\frac{1}{3}$.

Treating the bubble expansion as an isothermal process with $k=1$, transforms (\ref{eq:equilibrium_general_k}) to a cubic equation for $R_E\equiv Q_E^\frac{2}{5}$. When $\Delta p \in (-\Delta p_c,\Delta p_c)$ the solution is:
\begin{align}
    R_E=\frac{4S}{3\Delta p}\left \{\frac{1}{2}+\cos\left[\frac{1}{3}\arccos\left(1-2\left(\frac{\Delta p}{\Delta p_c} \right)^2\right) 
    \right. \right.
    \nonumber \\
    \left.\left.
    -\frac{2\pi n_\star}{3}\vphantom{\left(\frac{\Delta p}{\Delta p_c} \right)^2}\right]\vphantom{\left(\frac{\Delta p}{\Delta p_c} \right)^2}\right\}
    \label{eq:equilibrium-cubic-solution}
\end{align}
where $n_\star$ is chosen such that $R_E$ is the smallest positive root. The full solution is shown in Appendix \ref{appendix:solution-to-the-cubic}. 

Strong oscillations of the bubble size $\left(\max{Q} \gg Q_{E0}\right)$ thus require that at least at some part of the driving period $\Delta p(t) > \Delta p_c$.

Before concluding this section, we note that approximations which are often cited in the literature are written in the somewhat unnatural form: $R(t)=f(t)^\frac{2}{5}$ \cite{Putterman1995,Hunter1960,Obreschkow2012,Obreschkow2024}. This arbitrarily looking form in the $R$ frame has a clear meaning in the $Q$ frame. The assumption $R(t)=(A+vt)^\frac{2}{5}$ is nothing but a free motion of the particle in the $Q$ frame ($\dot{Q}=\text{const}$) whereas $R(t)=(A+at^2)^\frac{2}{5}$ corresponds to a motion with constant acceleration ($\ddot{Q}=\text{const}$).
\section{The Acoustic Cycle\label{sec:the-acoustic-cycle}}
\begin{figure*}[t] 
    \centering
    \begin{subfigure}[t]{0.45\textwidth}
        \centering
        \includegraphics[width=\linewidth]{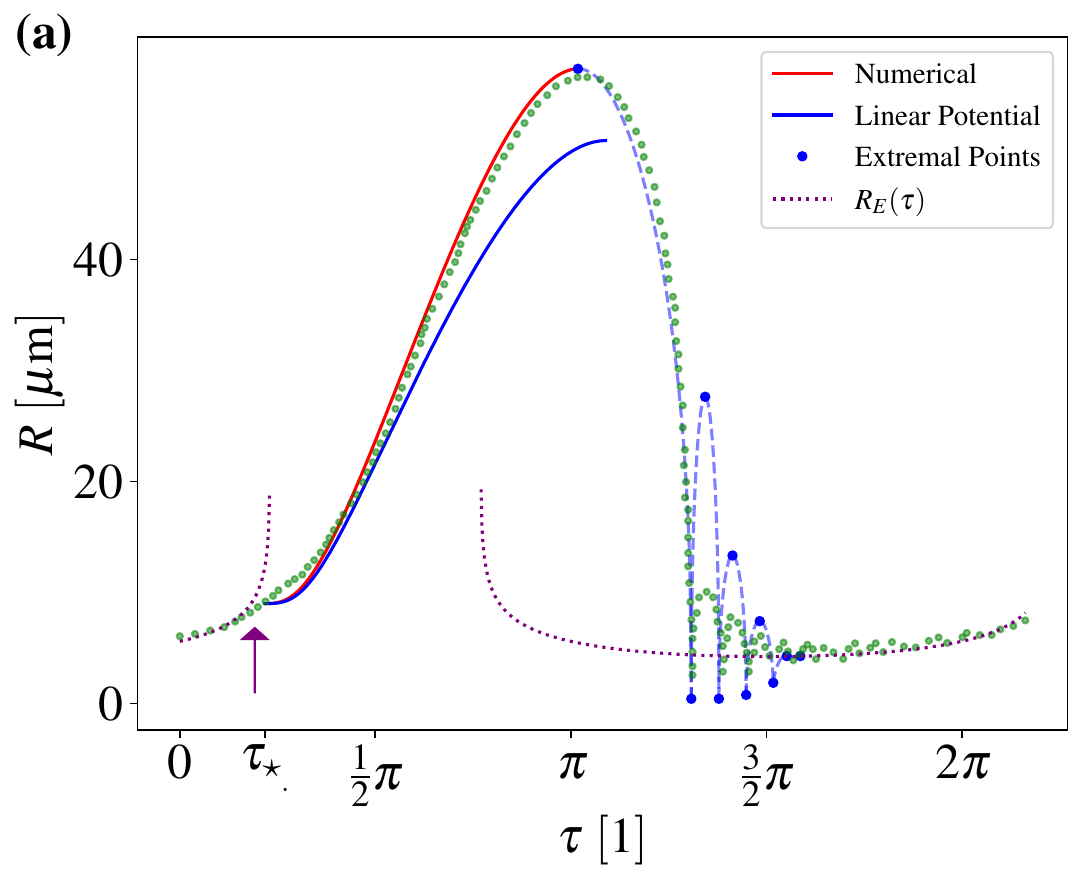} 
        \captionsetup[sub]{position=top}
    \end{subfigure}
    \hfill
    \begin{subfigure}[t]{0.45\textwidth}
        \centering
        \includegraphics[width=\linewidth]{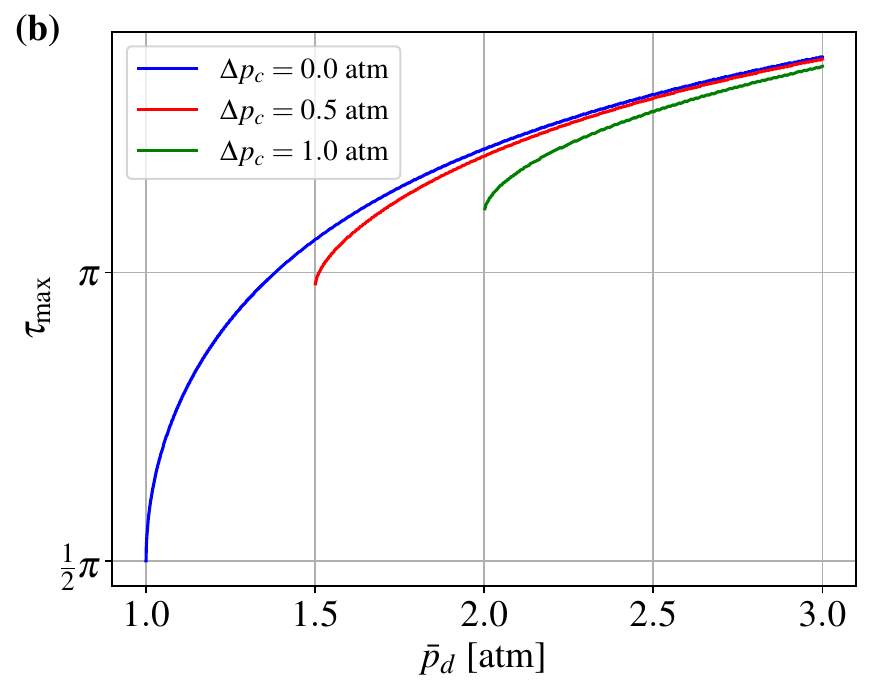} 
    \end{subfigure}
    \caption{\textbf{(a)} The bubble radius versus dimensionless time $\tau=\omega_dt$ ;  The green dots are experimental data from Putterman and Weninger, Annu. Rev. Fluid. 32, 445 (2000) \cite{Putterman2000}; licensed under a CC BY license. Initially and as long as the equilibrium radius changes slowly (between $\tau=0$ and the purple arrow), the bubble follows its equilibrium radius, $R_E(\tau)$, which is given by (\ref{eq:equilibrium-cubic-solution}). The blue curve represents the bubble expansion under the influence of a linear potential (Eq. (\ref{eq:q_in_far_from_equilibrium})) starting from $\tau_\star$, where the potential becomes non-binding. The red curve is obtained by numerically applying (\ref{eq:q_in_far_from_equilibrium}) to small time segments, adding the effect of the change in $Q$.  The collapse and after-bounces extremal points (the blue dots) were calculated using an iterative energy calculation outlined in Appendix \ref{appendix:collapse-extrema} (with $M=-0.9$). \textbf{(b)} The phase at which the bubble reaches its maximal size, $\tau_{\text{max}}$, versus the driving amplitude, $\bar{p}_d$. The figure assumes the limit of strong bubble oscillations $\left(\bar{p}_d\ge\Delta p_c+(p_0-p_V)\right)$ and negligible velocity at $\tau_\star$. If $\tau_{\max} \ge \pi$ Bjerknes force reversal is expected to be important, see Section \ref{sec:bjerknes_force_reversal}.}
    \label{fig:bubble-motion}
\end{figure*}
In this section, we present an analytical description of bubble motion over one acoustic cycle by dividing it into four segments: expansion near and far from the ambient equilibrium point, \( Q_{E0} \), collapse, and after-bounces.

We work in the driving time scale, $\tau=\omega_dt$, setting the bubble's initial conditions to $Q(\tau=0)=Q_{E0}$ with some positive velocity and assuming that the driving takes the form $p_d(\tau)=-\bar{p}_d \sin(\tau)$.

As $\tau$ increases, the driving goes negative, raising the tension, $\Delta p(t)$, and the instantaneous equilibrium point begins to increase slowly, $\frac{dQ_E(t)}{dt} / Q_E(t) \ll \omega(Q_E,t)$. Where $\omega(Q_E,t)$ is the frequency of oscillations around the minimum of the potential (see Eq. (\ref{eq:q_frequency_with_changing_expansion_point})). At this stage, the bubble follows the instantaneous equilibrium point and $Q(\tau)\approx Q_E(\tau)$.  For isothermal processes, with $k=1$, $Q_E(\tau)$ is analytically described by Eq. (\ref{eq:equilibrium-cubic-solution}). It is plotted as a dotted purple curve in Fig. \ref{fig:bubble-motion}a. At some moment, the change in the equilibrium point is too fast, i.e., $\frac{dQ_E(t)}{dt} / Q_E(t) \sim \omega(Q_E,t)$, and the bubble cannot follow it anymore. This instance is marked by the purple arrow in Fig. \ref{fig:bubble-motion}a.

Then, at $\tau_\star=\arcsin{\left(\frac{\Delta p_c+(p_0-p_V)}{\bar{p}_d}\right)}$, the tension is equal to the tensile strength, $\Delta p(\tau_\star) =\Delta p_c$, and the bubble equilibrium radius ceases to exist. The potential is non-binding and as long as $\Delta p(\tau) > \Delta p_c$ the bubble expands. When $Q \gg Q_{E0}$, the ideal gas pressure and surface tension are minute compared to the tension term and the potential can be assumed to take the form $U_k(Q,t)\sim \Delta p(t) Q^\frac{6}{5}$. In this limit, we may get a good description of the dynamics if we approximate the potential to be linear (see Fig. \ref{fig:incompressible-potential-around-a-point}). Under this approximation the force acting on the particle is $Q$-independent,  but time-dependent. The solution to the equation $\ddot{Q}=-\Delta p(\tau) Q(\tau_\star)^\frac{1}{5}$ may be expressed as a function of $\Delta\tau\equiv\tau-\tau_\star$:
\begin{align}
    Q(\tau)=Q(\tau_\star) + \alpha \Delta\tau^2 &+ \beta \Delta\tau 
    \nonumber \\
    &+\gamma \left[\sin(\tau_\star+\Delta\tau)-\sin(\tau_\star)\right]     \label{eq:q_in_far_from_equilibrium}
\end{align}
where $\alpha,\beta,\gamma$ are determined by the value of $\tau_\star, Q(\tau_\star)$ and $\frac{dQ}{d\tau}\Bigr|_{\substack{\tau=\tau_\star}}$ (the full form is shown in Eq. (\ref{eq:full_q_in_far_from_equilibrium})). Taking $Q(\tau_\star)$ from the experimental data \cite{Putterman2000} and neglecting $\frac{dQ}{d\tau}\Bigr|_{\substack{\tau=\tau_\star}}$ the blue curve in Fig. \ref{fig:bubble-motion}a was obtained. The linear potential approximation demonstrates good accuracy in determining the phase in which the bubble reaches its maximal radius, $\tau_{\text{max}}$. This phase plays a major role in the reversal of the Bjerknes force \cite{Leighton1990,Crum1975}. In Section \ref{sec:bjerknes_force_reversal} we utilize the analysis presented here to formulate a physical condition for the reversal of the Bjerknes force. Additionally, Fig. \ref{fig:bubble-motion}b shows $\tau_{\text{max}}$ as a function of the driving pressure amplitude, $\bar{p}_d$, for various tensile strengths, $\Delta p _c$. The figure assumes the limit of strong bubble oscillations and negligible velocity at $\tau_\star$. 

At $\tau_{\text{max}}$ the bubble reaches its maximum size and is ready to begin its collapse. From this instance until the end of the acoustic cycle, the tension remains below the tensile strength, $\Delta p(\tau) < \Delta p_c$. The potential then becomes binding, and the bubble oscillates around its minimum. During the first oscillation, the bubble undergoes its primary collapse, reaching sonic velocities and experiencing significant damping due to the compressibility of the liquid. After the primary collapse, the bubble quickly rebounds, resulting in rapid, damped oscillations around the equilibrium point. These oscillations occur on a fast time scale and are known as "after bounces".

In Section \ref{sec:the_compressible_regime} we develop a formula for the acoustic energy dissipation during the bubble's sonic collapse (Eq. (\ref{eq:acoustic_energy_dissipation_formula})). Using this formula, the after-bounces extremal points were calculated and plotted as blue points in Fig. \ref{fig:bubble-motion}a. Experimental data shows that around 99\% of the energy is lost between the first and second maximum points of the oscillations. Acoustic dissipation that is included in our model accounts for around 90\% of the energy loss, with the remaining energy loss accounted for by viscosity and luminescence \cite{Putterman2000}, which are not included in the model.
\section{Criterion for the Reversal of the Bjerknes Force \label{sec:bjerknes_force_reversal}}
So far, we have examined the oscillation of a driven bubble by considering a time-dependent acoustic drive, \( p_d = p_d(t) \). However, in reality, the acoustic drive also depends on position, \( p_d = p_d(t, z) \).

The space-dependent drive creates a pressure gradient $\nabla p_d$ that, in turn, exert a translational force on the bubble. The primary Bjerknes force is defined as the time-averaged force over one period of the drive, given by $F_B(z) = \langle -R^3(t)\nabla p_d(t, z)\rangle$ \cite{Bjerknes1906,Leighton1990,Crum1975,Louisnard2008}. For a standing acoustic wave, $p_d(t,z)=-\bar{p}_d \cos(kz)\sin(\omega_d t)$, the Bjerknes force is proportional to $\Phi$, the time correlator: 
\begin{align}
    \Phi \equiv -\frac{1}{2\pi R_{E0}^3}\int^{2
    \pi}_0{R^3(\tau)\sin{\left(
    \tau\right)}d\tau}
    \label{eq:bjerknes-force-reversal}
\end{align} where $
\tau=\omega_dt$. If $\Phi<0$, the bubble will move toward the antinodes of the driving, while if $\Phi > 0$, it will move toward the nodes. The value of $\Phi$ is determined by the relative phase between $R(t)$ and the drive $p_d(t)$.

In the weak drive regime, $p_d \ll p_0$, the tension is negative $\Delta p <0$ and the bubble can be approximated to oscillate in a simple harmonic potential around its ambient equilibrium radius, $R_{E0}$. The relative phase between the drive and the oscillation is either zero or $\pi$, depending on the relation between the bubble's internal frequency and the drive frequency \cite{Leighton1990}. If the drive frequency is higher than the internal frequency, the bubble will oscillate out-of-phase with the drive and vice versa. 

The bubble's internal frequency is inversely proportional to its equilibrium radius (see Eq. (\ref{eq:resonance-frequency-in-two-limits})). Thus, a critical radius $R_C$ can be defined such that bubbles with a small equilibrium radius, $R_{E0}<R_C$ will oscillate out-of-phase with the drive and vice versa. The sign of $\Phi$ will then be a function of $R_{E0}$ and $R_C$ only.

The physical picture for \textit{weakly} driven bubbles is now clear. Small bubbles, $R_{E0}<R_C$, will be directed to the pressure antinodes whereas big bubbles, $R_{E0}>R_C$, will be directed to the pressure nodes. 

This simple description breaks down when the bubble motion deviates from the linear regime \cite{Akhatov1997,Matula1997}. As explained in Section \ref{sec:the_potential_in_the_incompressible}, when the bubble tension overcomes the tensile strength $\Delta p \ge \Delta p_c$ at some point in the acoustic cycle, strong oscillations are expected to occur.  We can then calculate the time at which a small bubble ($R_{E0}<R_C$) reaches its maximum radius, which we define in dimensionless units as $\tau_{\text{max}}=\omega_d t_{\text{max}}$. 

The shift of $\tau_{\text{max}}$ from $\frac{\pi}{2}$ toward $\pi$, as shown in Fig. \ref{fig:bubble-motion}b, indicates a deviation from the linear regime. For weak driving, the Bjerknes force increases with the driving amplitude $\bar{p}_d$. However, when the driving becomes strong enough to push $\tau_{\text{max}}$ beyond $\pi$, the positive part of the integrand in (\ref{eq:bjerknes-force-reversal}) increases, resulting in a decrease in the Bjerknes force.

The condition $\tau_{\text{max}} \ge \pi$ can thus serve as a criterion for the reversal of the Bjerknes force. In Section \ref{sec:the-acoustic-cycle}, we examined the bubble's expansion. The instance when the tension, $\Delta p(\tau)$, equals the tensile strength was defined as $\tau_\star$. The subsequent expansion of the bubble was approximated by Eq. (\ref{eq:q_in_far_from_equilibrium}). Differentiating it with respect to time leads to an implicit equation for the bubble's extremal points (see Eq. (\ref{eq:implicit_extrema_formula})).

Drawing from the straightforward geometric argument provided in Appendix \ref{appendix:tau_max}, the condition $\tau_{\text{max}} \ge \pi$ can be approximated by a relation that depends solely on the tensile strength $\Delta p_c$ and the pressures:
\begin{align}
\tau_\star \ge \pi - \frac{\bar{p}_d}{p_0 - p_V}\left(1 + \cos(\tau_\star)\right)
\label{eq:tau-max-equivalent-condition}
\end{align}
where $\tau_\star = \arcsin{\left(\frac{\Delta p_c + (p_0 - p_V)}{\bar{p}_d}\right)}$. 
\section{The Compressible Liquid Regime\label{sec:the_compressible_regime}}
The compressibility of the liquid is important when the bubble undergoes its violent collapse \cite{Putterman2000,Putterman1995,Kamath1993,Lfstedt1993,Brenner2002,Prosperetti1986}. To account for it, the fairly complex Rayleigh-Plesset equation (\ref{eq:rayleigh_plesset}) has to be replaced by the even more complex \textit{general} Keller-Herring equation, valid to first order in $M$ \cite{Prosperetti1986}:
\begin{align}
	(1-(\lambda+1)M)R\ddot{R}+\left(1-\frac{1}{3}(3\lambda+1)M
	\right)\frac{3}{2}\dot{R}^2
	\label{eq:general_keller_herring}
	\nonumber \\ 
	= (1+(1-\lambda)M)h_B
	+ \frac{R}{c} \dot{h}_B
\end{align}
where $c$ is the speed of sound in the liquid,
\begin{align*}
    h_B \equiv \frac{p_{GE}\left(\frac{R_{E0}}{R}\right)^{3k}+\Delta p(t)-\frac{2S}{R}-\frac{4\nu_L\rho_L}{R}\dot{R}}{\rho_L}
\end{align*}  and $\lambda$ is a positive, "tunable" parameter that must not exceed $\frac{1}{M}$; for $\lambda=0$ equation (\ref{eq:general_keller_herring}) reduces to Keller's form \cite{Keller1956,Epstein1972,Keller1980}, and for $\lambda=1$ it reduces to Herring's form \cite{Herring1941,Trilling1952}. 

To extend our approach to the compressible regime, we modify the original transformation to:
\begin{align}
    X=R^{\alpha(M)} 
    \label{eq:compressible-liquid-transformation}
\end{align}
where 
\begin{align*}
    \alpha (M)\equiv \frac{5}{2}+\frac{M}{1-(\lambda+1)M}
\end{align*}
The kinetic energy in this case will be proportional to $R^D\dot{R}^2$, where $D$ is the problem's \textit{effective dimension}:
\begin{align}
	D(M) = 3 +\frac{2M}{1-(\lambda+1)M}
    \label{eq:effective-dimension}
\end{align}
The compressibility of the liquid then changes the dimension of the Rayleigh-Plesset equation, which is embedded in the coefficients of the equation's kinetic terms \cite{Klotz2013,Kudryashov2015,Wang2018}. 

In Appendix \ref{appendix:compressible-liquid-limit} we apply (\ref{eq:compressible-liquid-transformation}) to extend Eq. (\ref{eq:a_particle_in_a_potential}) to the compressible liquid regime (Eq. (\ref{eq:extended_q_energy})). In the limit of slowly varying Mach number, $\left(\frac{1}{\omega_d}\dot{M}, \frac{1}{\omega_d^2} \ddot{M} \ll 1\right)$ the product of the bubble's energy $E$ (Eq. (\ref{eq:bubble_energy_in_the_compressible_regime})) and its radius raised to the power $D(M)-3$ is conserved, up to viscosity losses, for time-independent external pressure. This limit can be referred to as the \textit{adiabatic limit}. The derivation can be found in Appendix \ref{appendix:compressible-liquid-limit}.

Neglecting viscosity and solving $\frac{d}{dt}\left(ER^{D(M)-3}\right)=0$ for $E$ yields:
\begin{align}
	E(R)=
    \begin{cases}
      E_0\left(\frac{R}{R_0}\right)^{3-D}  &  \dot{R}<0 \\
      E_0\left(\frac{R_0}{R}\vphantom{\frac{R}{R_i}}\right)^{D-3} &   \dot{R}>0
    \end{cases}
	\label{eq:energy_loss_for_collapse}
\end{align} 
where $E(t=0)=E_0$ and $R(t=0)=R_0$.

Equation (\ref{eq:energy_loss_for_collapse}), which is valid only when $M$ is nearly constant, i.e., when the velocity is nearly constant, leads to a remarkable observation: the energy, $E$, decreases both during expansion and during contraction of the bubble. This is because, during expansion the effective dimension $D(M)>3$, while, during contraction, $D(M)<3$.

We can use the considerations above to obtain a rough estimate of the energy loss to acoustic waves as the bubble contracts. We consider an oscillation that starts from the radius being $R_{\text{max}}$, passes through the equilibrium radius $R_E$, and ends at the minimal radius $R_{\text{min}}$. Most of the energy loss occurs in the regions of large $M$, which is between $R_E$ and $R_{\text{min}}$. We assume $M$ to be constant in that region, and obtain the following estimate:
\begin{align}
	E_{\text{acoustic}}=\frac{4}{3}\pi \left(p_0 - p_V \right) R^3_{\text{max}}
 \left(1-\left(\frac{R_{\text{min}}}{R_{E}}\right)^{3-D}\right)
 \label{eq:acoustic_energy_dissipation_formula}
\end{align} 
where we have assumed that the initial energy is $\frac{4}{3}\pi \left(p_0-p_V\right)R_{\text{max}}^3$. It is possible to deduce an approximated formula for either $R_{\text{max}}$ as a function of $R_{\text{min}}$ and $R_E$ or $R_{\text{min}}$ as a function of $R_{\text{max}}$ and $R_E$ by assuming conservation of energy at zeroth order in $M$ (Eq. (\ref{eq:min-from-max})). 

By applying (\ref{eq:acoustic_energy_dissipation_formula}) with $M=-0.9$ and $\lambda=0$, for which the effective dimension is approximately $2.05$, the after-bounces of the bubble were calculated with good accuracy. The result is shown in Fig. \ref{fig:bubble-motion}a. The calculation procedure is outlined in Appendix \ref{appendix:collapse-extrema}.
\section{Summary}
We have introduced a novel perspective on the highly nonlinear dynamics of a driven bubble. In the incompressible liquid regime, we developed an exact description (Eq. (\ref{eq:a_particle_in_a_potential})) that is analogous to the dynamics of a Newtonian particle in a potential. Within the compressible liquid regime, our understanding is limited to an approximate model that relies on an adiabatic treatment of the bubble's Mach number (Eq. (\ref{eq:energy_loss_for_collapse})). We have demonstrated how these frameworks can be used to analytically calculate various aspects of the bubble's dynamics.
\appendix
\section{The Equilibrium Radius in the Isothermal Case\label{appendix:solution-to-the-cubic}}
In Section \ref{sec:the_potential_in_the_incompressible} we derived an implicit equation for the extremal points of the bubble, as shown in Eq. (\ref{eq:equilibrium_general_k}). In the isothermal case, where $k=1$, Eq. (\ref{eq:equilibrium_general_k}) simplifies to a cubic equation for $R_E$. In this appendix we show how the resulting cubic equation was \textit{analytically} solved.

The solution to a cubic equation can be found using Cardano's method when there is one real root \cite{Gindikin2007}, and Viète's trigonometric method when there are three real roots \cite{Nickalls2006}. The stable equilibrium radius is the smallest positive root.

We first transform (\ref{eq:equilibrium_general_k}) to a depressed polynomial for $\sigma(t)=R_E(t) - \frac{2S}{3(p_v-p_\infty(t))}$. Then the discriminant is proportional to
\begin{align}
    \Delta(t) = \frac{Q_{E0}^\frac{12}{5}p^2_{GE}(\Delta p-\Delta p_c)(\Delta p+\Delta p_c)}{\Delta p^4}
\end{align}
When $-\Delta p_c<\Delta p<\Delta p_c$ the discriminant is negative and there are three real roots. In this scenario, Viète's trigonometrical method is applied. The solution was shown in Eq. (\ref{eq:equilibrium-cubic-solution}). Conversely, if the discriminant is positive, there is one real root and Cardano method is applied:
\begin{align}
\sigma(t)=\sqrt[3]{-\frac{q}{2}+\sqrt{\Delta}}+\sqrt[3]{-\frac{q}{2}-\sqrt{\Delta}}
\label{eq:cardano-method}
\end{align}
where $q=\frac{-2^4S^3+27\Delta p^2p_{GE}Q_{E0}^\frac{6}{5}}{27\Delta p^3}$. Note that when $\Delta p > \Delta p_c$, (\ref{eq:cardano-method}) yields a negative root. This root does not correspond to a \textit{physical} equilibrium radius. 
\section{Harmonic Potential Approximation\label{appendix:harmonic-approximation}}
In this appendix we derive the harmonic approximation of the bubble's potential in the incompressible liquid regime (Eq. (\ref{eq:the_incompressible_potential})) around a point and explicitly write the solution for the bubble motion under a linear but time-dependent potential.

\begin{figure}
    \centering
    \includegraphics[width=1\linewidth]{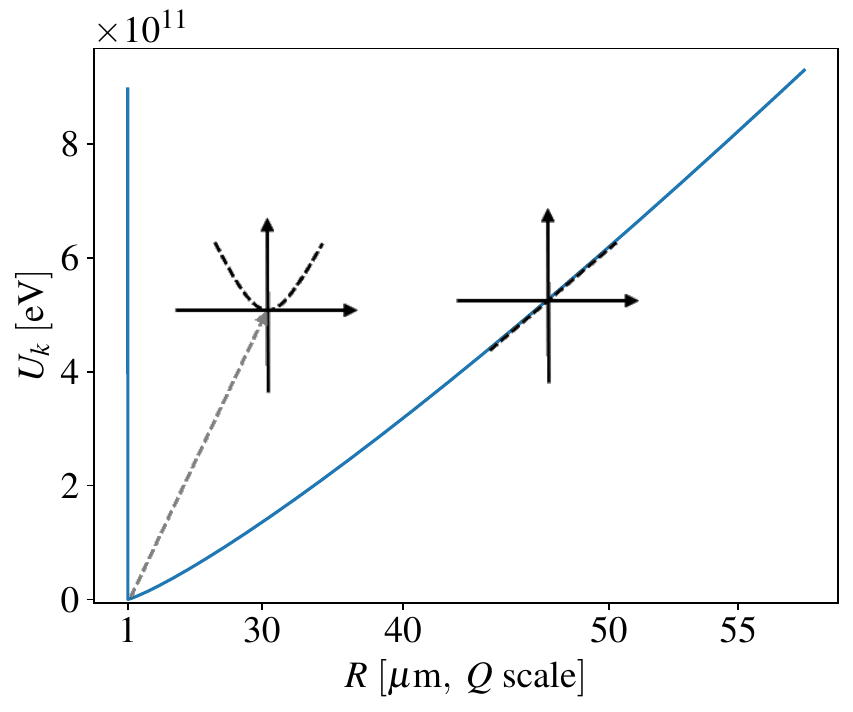}
    \caption{Expansion of the bubble potential around a point, $Q_0$. Assuming $Q=Q_0+q$, the perturbation, $q$, is evolving according to $\ddot{q}=-\left(\frac{dU_k}{dq}\Big|_{q=Q_0} + \frac{d^2U_k}{dq^2}\Big|_{q=Q_0} q\right)$. If the expansion is done far from the equilibrium point, the effective potential is approximately linear at short distances.}
    \label{fig:incompressible-potential-around-a-point}
\end{figure}

Working in the driving time scale $\tau=\omega_d t$ where $p_d(\tau)=-\bar{p}_d\sin(\tau)$, denoting $Q=Q_0+q$ and expanding $U_k$ to second order in $q$ (as illustrated in Fig. \ref{fig:incompressible-potential-around-a-point}) yields the linear equation:
\begin{align}
    \frac{d^2q}{d\tau^2}
    =\frac{5}{2} \frac{Q_0^\frac{1}{5}}{\omega_d^2}
   h_B(Q_0)
    -\left(\frac{\omega(Q_0,\tau)}{\omega_d}\right)^2q
    \label{eq:harmonic_oscillator_in_q}
\end{align}
where $ h_B(Q) \equiv \frac{p_{GE}\left(\frac{Q_{E0}}{Q}\right)^{\frac{6k}{5}}+\Delta p(t)-\frac{2S}{Q^\frac{2}{5}}}{\rho_L}$ is the bubble enthalpy and we have neglected viscosity. The bubble's internal frequency is:
\begin{align}    
    \omega^2(Q_0,t)=
    \frac{1}{2\rho_LQ_{0}^{\frac{4}{5}}}
    \left(
        \left[
            (6k-1)\left(\frac{Q_{E0}}  {Q_0}\right)^\frac{6k}{5}+1
            \right](p_0-p_V)
            \right.\nonumber \\
            +p_d(t)
            +\frac{2S}{Q_0^\frac{2}{5}}\left[
                (6k-1)\left(\frac{Q_{E0}}         {Q_0}\right)^\frac{6k}{5}-1
            \right]
            \left. \vphantom{\left(\frac{Q_{E0}}{Q_0}\right)^\frac{6k}{5}}\right)
    \label{eq:q_frequency_with_changing_expansion_point}
\end{align} where $Q_{E0}$ is the equilibrium point in the absence of the drive (referred to as the "ambient equilibrium point")

Two limiting cases can be examined, one in which the expansion point is very far from the ambient equilibrium point, $Q_0 \gg Q_{E0}$ and the second in which the two are comparable $Q_0 \sim Q_{E0}$:
\begin{align}
	\omega^2 \sim \frac{1}{2\rho_LQ_0^{\frac{4}{5}}}
    \begin{cases}
     6k(p_0-p_V)+p_d(t)+\frac{4(3k-1)S}{Q_0^{\frac{2}{5}}} &  Q_0\sim Q_{E0} \\
      -\Delta p(t) &   Q_0 \gg Q_{E0}
    \end{cases}
	\label{eq:resonance-frequency-in-two-limits}
\end{align}  where surface tension was neglected for $Q_0 \gg Q_{E0}$. 

The factor of $6k$ in front of the ambient and vapor pressures in the case of $Q_0 \sim Q_{E0}$ means that the driving pressure weakly affect the bubble's internal frequency around the ambient equilibrium point. This factor is absent in the case of $Q_0\gg Q_{E0}$. 

Eq. (\ref{eq:harmonic_oscillator_in_q}) can be reduced to the linear potential case by assuming that $\omega \ll \omega_d$. Imposing initial conditions at $\tau_\star$, $Q(\tau_\star)=Q_\star$ and $\frac{d}{d\tau}Q(\tau_\star)=v_\star$ an analytical solution can be written in terms of $\Delta \tau \equiv \tau-\tau_\star$:
\begin{align}
    q / v_0
    =&-\frac{\Delta\tau^2}{2}
    +\left(\frac{v_\star}{v_0}+\frac{\bar{p}_d / \rho_L}{\left|h_B(Q_\star)+\frac{p_d}{\rho_L}\right|}\cos{(\tau_\star)}\right)\Delta\tau
    \nonumber  \\
    -&\frac{\bar{p}_d / \rho_L}{\left|h_B(Q_\star)+\frac{p_d}{\rho_L}\right|}(\sin(\tau_\star +\Delta\tau)-\sin(\tau_\star))
    \label{eq:full_q_in_far_from_equilibrium}
\end{align}
where $v_0=\frac{5}{2} Q_\star^\frac{1}{5}\left|h_B(Q_\star)+\frac{p_d}{\rho_L}\right| / \omega_d^2$. The plus sign in front of $\frac{p_d}{\rho_L}$ results from the subtraction of $-\frac{p_d}{\rho_L}$ from the bubble’s enthalpy $h_B(Q_\star)$. It is crucial to acknowledge that due to the omission of dissipation in deriving Eq. (\ref{eq:full_q_in_far_from_equilibrium}), this equation does not represent a steady-state solution.
\section{Applying the Transformation in the Compressible Liquid Regime\label{appendix:compressible-liquid-limit}}
By applying (\ref{eq:compressible-liquid-transformation}) on the general Keller-Herring equation (Eq. (\ref{eq:general_keller_herring})), integrating and dropping $O(M^2)$ terms, an energy balance equation can be written as:
    \begin{align}
	&\frac{d}{dt}
\left(\rho_L\dot{X}^2+\left(U_k(X)+MU_M(X)\right)X^{\frac{D-3}{\alpha}} \right)
        =
    \nonumber \\
 &\frac{5}{12}\left(5-\frac{13}{3}M\right)\dot{p}_dX^\frac{D}{\alpha}
 -4\left(1+ M\right)\rho_L\nu_L \frac{\dot{X}^2}        {X^{\frac{2}
 {\alpha}}} 
	-f(\dot{M},\ddot{M})
 \label{eq:extended_q_energy}
\end{align}
where $U_k(X)$ is the potential in the incompressible liquid case, defined in Eq. (\ref{eq:the_incompressible_potential}) and for $k\ne1$:
\begin{align}
    U_M(X) =-\frac{40}{9}\Delta p(t)X^{\frac{3}{\alpha}}
    +\mu p_{GE}R_{E0}^{3k}X^{\frac{3-3k}{\alpha}}
    +5SX^{\frac{2}{\alpha}} 
\end{align}
wherein $\mu= -\frac{5}{36}\frac{45k^2-87k+32}{(k-1)^2}$. 

In the absence of the periodic driving the first term of the right-hand side vanishes. Then, for $M=0$ the time derivative of the energy (the left-hand side) is the energy dissipated to viscosity (the second term in the right-hand side), just as expected from the motion of a damped particle in a potential. When $M\ll 1$ and is slowly varying $\left(\frac{1}{\omega_d}\dot{M}, \frac{1}{\omega_d^2} \ddot{M} \ll 1\right)$, the third term of the right-hand side, $f(\dot{M},\ddot{M})$, may be neglected. Then, the left-hand side is a modified expression for a quantity that is conserved up to viscosity losses. 

In this limit, which can be referred to as the adiabatic limit, it is instructive to come back to the bubble radius, $R$. The conserved quantity (the left-hand side of Eq. (\ref{eq:extended_q_energy})) may be written, in that limit, as the product of the bubble's energy, $E$, and $R^{D(M)-3}$, where the energy is 
\begin{align}
    E=\rho_LR^3\dot{R}^2+U_k(R)+MU_M(R)
    \label{eq:bubble_energy_in_the_compressible_regime}
\end{align}and by neglecting viscosity losses, we get
\begin{align}
	\frac{d}{dt}\left(ER^{D(M)-3}\right)
    =0
	\label{eq:adiabatic_limit_integrable}
\end{align} 
as in Section \ref{sec:the_compressible_regime} of the main text.
\section{Collapse Extremal Points Calculation\label{appendix:collapse-extrema}}
The collapse and subsequent after-bounces extremal points are plotted as blue dots in Fig. \ref{fig:bubble-motion}a. In this appendix, we outline the method for calculating these extremal points. 

First, we denote the sequence of maximal and minimal radii as \( R_{\text{max}}^{(n)}\) and \(R_{\text{min}}^{(n)} \), respectively. At the beginning of the calculation, the initial maximal radius, \( R_{\text{max}}^{(0)} \), and the equilibrium radius, \( R_{E0} \), must be specified. Then, the initial energy is taken to be $E^{(0)}=\frac{4\pi}{3} \left(R_{\text{max}}^{(0)}\right)^3(p_0-p_V)$. 

Next, by assuming conservation of energy to zeroth order in \( M \), the sequential minimal radius, \( R_{\text{min}}^{(0)} \), can be determined from Eq. (\ref{eq:the_incompressible_potential}):
\begin{align}
    R_{\text{min}}^{(n)}=\left(\left(\frac{1}{k-1}\frac{p_{GE}}{p_0-p_v}\right)^\frac{1}{3}\frac{R_{E0}^{k}}{R_{\text{max}}^{(n)}}\right)^\frac{1}{k-1}
    \label{eq:min-from-max}
\end{align} where we neglected surface tension and assumed that at $R_{\text{max}}^{(n)}$ the internal pressure term, $p_{GE}$, is negligible and at $R_{\text{min}}^{(n)}$ the tension term is negligible. The minimal radius was also constrained to be at least the van der Waals core radius.

After calculating $R_{\text{min}}^{(n)}$, the subsequent maximal radius $R_{\text{max}}^{(n+1)}$ must be determined. It is assumed that the bubble loses energy during its motion from its equilibrium radius to the minimal radius, as dictated by Eq. (\ref{eq:acoustic_energy_dissipation_formula}). This allows us to calculate $E^{(n+1)}$ from $E^{(n)}$ by specifying $M$. The sequential maximal radius is now obtained from energy conservation:
\begin{align}
    R_{\text{max}}^{(n+1)}=R_{\text{max}}^{(n)} \sqrt[3]{\left(\frac{R_{\text{min}}^{(n)}}{R_{E0}}\right)^{3-D(M)}}
\end{align}
and we are ready to repeat the calculation. 
The temporal spacing between the first max and min radii was taken from the experimental data. The temporal spacing between the other extremal points was obtained from the bubble internal frequency (Eq. (\ref{eq:q_frequency_with_changing_expansion_point})).
\begin{figure}[t]
    \centering
    \includegraphics[width=1\linewidth]{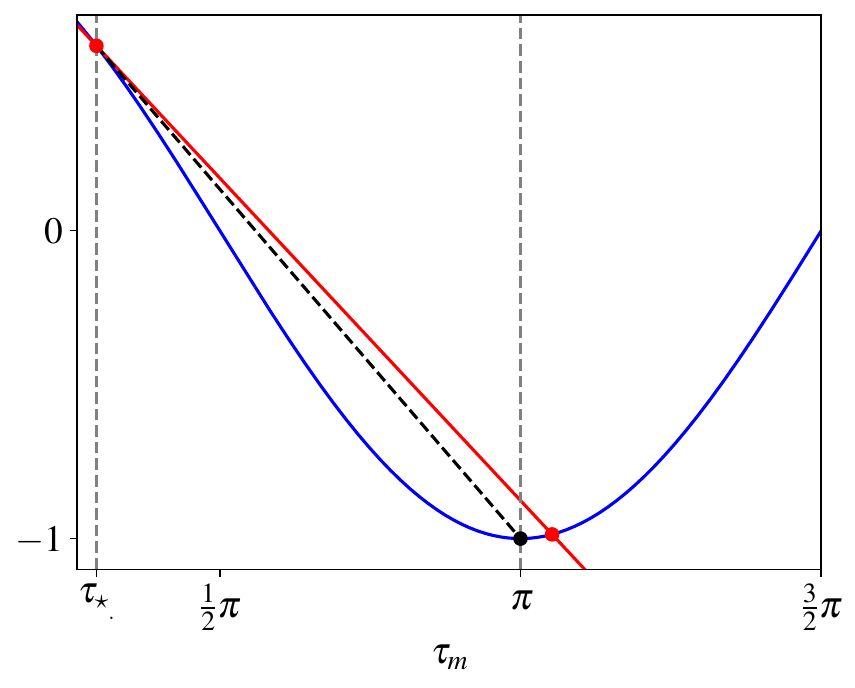}
    \caption{Visualization of the geometric argument of Appendix \ref{appendix:tau_max}. The blue curve is the left-hand side of Eq. (\ref{eq:implicit_extrema_formula}) whereas the red curve is the right hand side. The red intersection points are the solutions. The second red intersection corresponds to $\tau_{\text{max}}$. If the slope of the red curve exceeds that of the linear curve connecting the points $(\tau_\star, \cos\tau_\star)$ and $(\pi, -1)$ (represented by the black dashed line), then $\tau_{\text{max}} > \pi$, and vice-versa.}
    \label{fig:tau_max_condition}
\end{figure}
\section{Geometrical Argument for the Reversal of the Primary Bjerknes Force\label{appendix:tau_max}}
In Section \ref{sec:bjerknes_force_reversal}, we established that Bjerknes force reversal is expected when the time it takes for the bubble to reach its maximum radius, denoted in dimensionless units as $\tau_{\text{max}}=\omega_d t_{\text{max}}$, exceeds $\pi$. In this appendix, we explicitly derive Eq. (\ref{eq:tau-max-equivalent-condition}), which provides an approximation for the condition $\tau_{\text{max}} \geq \pi$.

First, we differentiate Eq. (\ref{eq:full_q_in_far_from_equilibrium}) with respect to $\tau=\omega_dt$ yielding an implicit equation for the phases at the bubble's extremal points, $\tau_m$:
\begin{align}
        \cos(\tau_m)=\cos(\tau_\star)-\left(\tau_m- \tau_\star\right) \frac{p_0-p_V}{\bar{p}_d}
    \label{eq:implicit_extrema_formula}
\end{align}
where $\tau_\star = \arcsin{\left(\frac{\Delta p_c + (p_0 - p_V)}{\bar{p}_d}\right)}$ and we approximated $\left|h_B(Q_\star) + \frac{p_d}{\rho_L}\right| \approx \frac{p_0 - p_V}{\rho_L}$ and assumed that the velocity at $\tau_\star$ is negligible, i.e., $\frac{v_\star}{v_0} \ll 1$.

Eq. (\ref{eq:implicit_extrema_formula}) has two solutions between $\tau_\star$ and $2\pi$. The first solution is $\tau_m=\tau_\star$ and the second solution is $\tau_m=\tau_{\text{max}}$. The right hand side is a \textit{linear} function of $\tau_{\text{m}}$ with $-\frac{p_0-p_V}{\bar{p}_d}$ being the slope. The condition $\tau_{\text{max}} \geq \pi$ is therefore equivalent to requiring that this slope will be greater or equal the slope of the linear curve connecting the points $(\tau_{\star}, \cos{\tau_{\star}})$ and $(\pi, -1)$:
\begin{align}
    -\frac{p_0-p_V}{\bar{p}_d} \ge \frac{-1-\cos{\tau_{\star}}}{\pi-\tau_\star}
\end{align}
which is equivalent to Eq. (\ref{eq:tau-max-equivalent-condition}). The argument is visualized in Fig. \ref{fig:tau_max_condition}.
\nocite{*}
\bibliography{paper/citations}

\end{document}